\documentclass[aps,reprint,prb,twocolumn]{revtex4-1}

\usepackage{amsmath}
\usepackage{amsfonts}
\usepackage{amssymb}
\usepackage{graphicx}
\usepackage{moreverb}
\usepackage{url}
\usepackage{epsf}
\usepackage{color}
\usepackage{multirow}
\usepackage{subfigure}
\usepackage{bm}

\begin{document}

\title{Calculating optical absorption spectra for large systems using linear-scaling density-functional theory}
\author{Laura E. Ratcliff}
\email[E-mail: ]{laura.ratcliff08@imperial.ac.uk}
\author{Nicholas D. M. Hine}
\author{Peter D. Haynes}
\affiliation{Department of Materials, Imperial College London, London SW7 2AZ, United Kingdom}
\date{\today}

\begin{abstract}
A new method for calculating optical absorption spectra within linear-scaling density-functional theory (LS-DFT) is presented, incorporating a scheme for optimizing a set of localized orbitals to
accurately represent unoccupied Kohn-Sham states. Three different schemes are compared and the most promising of these, based on the use of a projection operator, has been implemented in a fully-functional
LS-DFT code. The method has been applied to the calculation of optical absorption spectra for the metal-free phthalocyanine molecule and the conjugated polymer poly(\emph{para}-phenylene). Excellent agreement with
results from a traditional DFT code is obtained.
\end{abstract}

\maketitle

\section{Introduction}
\label{sec:intro}

Theoretical spectroscopy is a a tool of growing importance both in understanding experimental results and making predictions about new materials.  Using simulation, it is possible to analyse spectra to a level of detail which is hard to achieve experimentally, for example by identifying which electronic transitions correspond to a particular peak, or by observing the effect of small changes in the electronic structure on the optical spectra.  The information obtained can help with the interpretation of experimental results, or can be used in tandem with experiment to enable the development of materials with a particular property in mind.  

Density-functional theory (DFT)~\cite{hohenberg42,kohn43} is a good initial framework in which to calculate the energy eigenstates required for such spectra.  In practice, however, many systems of interest are large in scale, and as such computationally expensive, if not impossible, to treat with traditional approaches to DFT, where the computational effort scales as the cube of the system size.  However, DFT can also be reformulated to scale only linearly with system size, which requires the use of local orbitals~\cite{giulia_locorb,ordejon_locorb,hernandez_locorb,fatterbert_locorb,skylaris_ngwfs}.  This offers the opportunity to access much larger system sizes, and if combined with theoretical spectroscopy, it could become a very powerful tool.  To this end, a method has been developed for the calculation of optical absorption spectra within linear-scaling DFT methods, which tackles some of the challenges that arise due to the use of local orbitals.  It could also be extended to other types of spectroscopy in future.

Linear-scaling methods use local orbitals which are optimized to describe the occupied states.  There are two approaches to the optimization of such orbitals; either via the use of basis sets of purpose-designed atomic orbitals, or via the minimization of total energy with respect to some set of local orbitals which therefore become adapted to the system in question, which is the approach followed in this work.  In both cases, this results in a basis which is unable to represent the unoccupied states very well.  This problem is particularly noticeable in systematic linear-scaling methods such as \textsc{onetep}~\cite{onetep1,onetep2,onetep3,onetep_forces}, where the equivalence of the underlying basis with plane-wave methods means that after optimization of the local orbitals to minimize the total energy, the occupied states are in very precise agreement with plane wave results, but the unoccupied states may be significantly in error. Therefore, in this work a new method is presented whereby a second set of localized functions is optimized to describe the unoccupied states.  With this method, it becomes possible to implement the calculation of optical absorption spectra using Fermi's golden rule. 

Due to the inherent deficiencies in DFT, in particular the fact that there is no theoretical relation between the Kohn-Sham states and the true quasi-particle energies, this will of course only be an approximate method for the calculation of optical spectra.  However, in practice reasonable agreement has been seen with experiment, particularly when the scissor operator approximation~\cite{godby_sciss,bald_sciss} is employed.  Furthermore, as the emphasis within this work is on application to large systems, more accurate methods such as the \emph{GW} approximation~\cite{hedin64,godby65,Sole97} are prohibitively expensive, and so the approximation becomes justified with respect to the aims of studying previously inaccessible system sizes whilst maintaining a reasonable standard of accuracy.

The relevant methodology will be briefly outlined in the following section, highlighting in particular the reasons why the unoccupied states are not calculated to a high level of accuracy when the total energy is minimized.  The methodology for testing different conduction optimization methods and the results obtained will be described.  This will be followed by the relevant details of the implementation in \textsc{onetep}.  To conclude the methodological section, a description of the calculation of optical absorption spectra will be presented.  In section~\ref{sec:results}, results will be presented on both a molecular and an extended system: metal-free phthalocyanine and poly(\emph{para}-phenylene), followed by the conclusion in section~\ref{sec:conclusions}.

\section{Methodology}
\label{sec:method}

\subsection{Linear-scaling density-functional theory with local orbitals} \label{sec:lin_sc_dft}

It is well known that for quantum mechanical systems containing a large number of interacting particles, physical processes are usually only affected by their immediate locality, a fact which has been referred to as the principle of `nearsightedness'~\cite{kohn16}.  More precisely, it has been established that the single-particle density matrix will decay exponentially with respect to distance for systems with a band gap~\cite{ismail46,kohn45,he81}.  One therefore ought to be able to take advantage of this principle in order to develop linear-scaling formalisms of DFT, and indeed a variety of such methods exist, which have been the subject of various reviews~\cite{goedecker19,galli14,bowler_review}.  One such method is that employed in \textsc{onetep}, which has been discussed in detail elsewhere~\cite{onetep1,onetep2,onetep3} but for which the key points will now be summarized.

One of the features necessary for the development of a linear-scaling method is the use of localized basis functions; in the case of \textsc{onetep}, a set of non-orthogonal generalized Wannier functions (NGWFs)~\cite{skylaris_ngwfs} are used, which are atom-centered and strictly localized within a set radius.  These NGWFs are represented in terms of a basis set of periodic cardinal sine (psinc) functions~\cite{mostofi_prec}, which can be related to plane-waves, and are optimized during the calculation to create a minimal basis which is adapted specifically to reflect the chemical environment of the system in question.  This can be seen from the elimination of basis set superposition errors, which commonly occur in other approaches using localized basis sets~\cite{onetep_bsse}.

To avoid the need for orthogonalizing extended orbitals, a density matrix (DM) representation is adopted, rather than explicit wavefunctions.  The density operator is formally defined as:
\begin{equation} \label{eq:dens_op}
\hat{\rho}=\sum_nf_n|\psi_n\rangle\langle\psi_n|,
\end{equation}
where the $\{\psi_n\left(\mathbf{r}\right)\}$ are the Kohn-Sham orbitals, the $f_n$ are their occupation numbers and the density matrix, $\bm{\rho}$ is found from the density operator using ${\rho}_{\alpha\beta}=\langle\phi_{\alpha}|\hat{\rho}|\phi_{\beta}\rangle$.  For a non-orthogonal basis the density operator can equivalently be written in the following separable form~\cite{mcweeny41,hernandez_kernel}:
\begin{equation}
\hat{\rho}=\sum_{\alpha\beta}|\phi_{\alpha}\rangle{K^{\alpha\beta}}\langle\phi_{\beta}| ,
\end{equation}
where $K^{\alpha\beta}$ is the density kernel and $\{\phi_{\alpha}\left(\mathbf{r}\right)\}$ are the NGWFs.  In this form, when combined with the locality of the NGWFs, it becomes possible to truncate the density kernel.  The Hamiltonian, kernel and overlap matrices then become sparse and so can be multiplied together in order $N$ operations.  The DM is required to be idempotent, using a combination of the McWeeny purification transformation~\cite{mcweeny41} and penalty functionals~\cite{mcweeny41,haynes_penalty}.  

In this way \textsc{onetep} combines the high accuracy of plane-wave calculations via the use of a psinc basis set, with the speed of minimal basis approaches via the use of \emph{in-situ} optimized, localized NGWFs~\cite{skylaris_si}.  Furthermore the NGWF optimization process also allows for insight into the local chemical environment which is reflected in their final state.  \textsc{onetep} is particularly suited to lower dimensional systems, as empty space which is not covered by the atom-centered NGWFs is virtually free from the point of view of computational effort.  It should also be noted that \textsc{onetep} is designed for application to large systems, either with large unit cells or using the supercell approximation, so that only a single $\mathbf{k}$-point need be treated.  This is chosen to be the $\Gamma$-point, which has the added benefit that the Kohn-Sham eigenstates and therefore the basis set and related quantities can be chosen to be real.

In a standard \textsc{onetep} calculation the energy and density are determined from the DM and NGWFs, while the individual eigenstates are not explicitly considered.  They can, however, be recovered by a single diagonalization of the Hamiltonian matrix in the basis of NGWFs at the end of a calculation, but only the occupied Kohn-Sham orbitals are accurately represented.  This is because the NGWF optimization is solely focussed on minimizing the bandstructure energy of the occupied states, resulting in a basis that does not accurately represent the unoccupied states~\cite{onetep_dos_cond}.  In practice some of the lower lying conduction states are close to the correct values, particularly when they are of a similar character to the valence states, however conduction states which are higher in energy are poorly treated and some can be completely absent.  Therefore in order to correctly calculate densities of states, band structures and in particular spectra, where matrix elements between valence and conduction states are needed, it becomes necessary to consider the optimization of a second set of NGWFs.  

It should be noted here that various methods exist for calculating electronic excitation energies using the \emph{GW} method, which avoid the need for explicitly summing over unoccupied states in order to increase computational efficiency~\cite{gw_no_cond,gw_no_cond2,gw_no_cond3,gw_no_cond4}.  Whilst this would appear to invalidate the need for a method of accurately calculating the unoccupied states, it is still necessary to have a complete basis in order to define a projection operator onto the conduction manifold that requires the identity operator.  Therefore even with the existence of such approaches it is important to have a method of creating a basis which is able to accurately represent both the occupied and unoccupied states.

\subsection{Methods for calculating unoccupied states} \label{sec:cond_methods}

Possible methods for optimizing a new set of NGWFs to represent the conduction states include the folded spectrum method~\cite{folded_orig,folded}, the shift-invert method~\cite{shift_invert} and the use of a projection operator.  These differ principally by the form of the eigenvalue equation they attempt to solve to obtain the excited states.

A toy model was created within which these methods were compared.  It was required to imitate the main features of a systematic local-orbital method, whilst remaining as simple as possible.  This included the use of an iterative minimization scheme using conjugate gradients, with a preconditioning scheme equivalent to that used in \textsc{onetep}~\cite{mostofi_prec}, a range of localized basis sets, of which B-splines~\cite{bsplines} were found to be the most accurate, and simple one-dimensional potentials.

\begin{figure*}[]
    \begin{center}
    \includegraphics[scale=0.6]{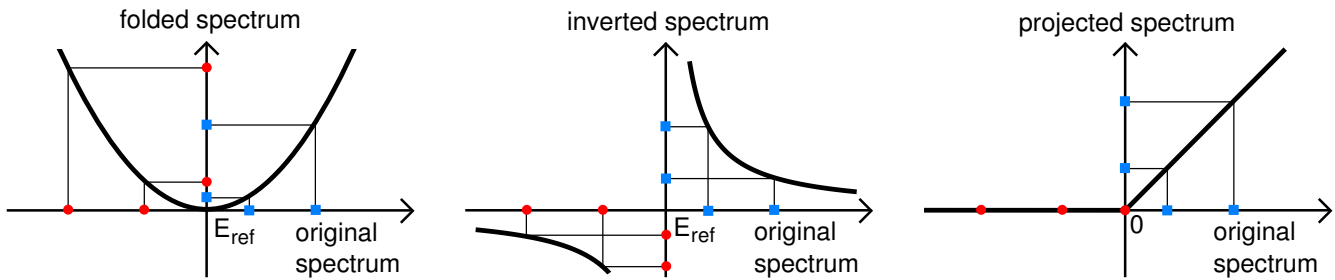}
  \caption{(Color online) Schematics comparing the three methods for the calculation of unoccupied states.  The original spectrum is shown on the $x$-axis and the transformed spectrum on the $y$-axis, with thick black curves depicting the relationship between the two sets of eigenvalues.  The occupied states are shown as (red) circles and the unoccupied states as (blue) squares, with the reference energy arbitrarily chosen to be in the gap for the folded spectrum and shift invert methods.}
  \label{fig:three_methods}
  \end{center}
\end{figure*}

\paragraph{Folded spectrum} \label{subsection:folded}

The folded spectrum method involves folding the energy spectrum of a matrix $\mathbf{H}$ around a reference energy $E_{\text{ref}}$, where the spectrum of $\mathbf{H}$ is found from the eigenvalue equation $\mathbf{H}\mathbf{x}=\epsilon\mathbf{x}$.  This leads to a new eigenvalue equation which the eigensolutions of the original equation also satisfy:
\begin{equation}
\left(\mathbf{H}-E_{\text{ref}}\mathbf{I}\right)^2\mathbf{x}=\left(\epsilon-E_{\text{ref}}\right)^2\mathbf{x} .
\end{equation}
The smallest eigenvalues of this new matrix are related to those of $\mathbf{H}$ nearest $E_{\text{ref}}$, so that by setting $E_{\text{ref}}$ to a value near the center of the energy range covered by the conduction states, they can be found by solving the new eigenvalue equation.  It can also be generalized to account for the use of a non-orthogonal basis set~\cite{gen_folded}, giving the following: 
\begin{equation}
\left(\mathbf{H}-E_{\text{ref}}\mathbf{S}\right)\mathbf{S}^{-1}\left(\mathbf{H}-E_{\text{ref}}\mathbf{S}\right)\mathbf{x}=\left(\epsilon-E_{\text{ref}}\right)^2\mathbf{S}\mathbf{x}.
\end{equation}
This is illustrated by Fig.~\ref{fig:three_methods}, which contains a schematic showing the effect of the folded spectrum method on a set of example eigenvalues.  This method has been used previously for example to study the conduction band minimum for silicon within the tight-binding method~\cite{martins79}, as well as in studies of quantum dots~\cite{folded_dot,folded_dot2}.

\paragraph{Shift invert} \label{subsection:shift}

Shift invert is another method of spectral transformation which can be used to find extremal eigenvalues. Starting from a given generalized eigenvalue equation $\mathbf{H}\mathbf{x}=\epsilon\mathbf{S}\mathbf{x}$, the Hamiltonian is shifted with respect to some reference energy and then inverted, giving:
\begin{equation}
\left(\mathbf{H}-E_{\text{ref}}\mathbf{S}\right)^{-1}\mathbf{S}\mathbf{x}=\left(\epsilon-E_{\text{ref}}\right)^{-1}\mathbf{x} .
\end{equation}
However, even if both $\mathbf{S}$ and $\mathbf{H}$ are Hermitian, $\left(\mathbf{H}-E_{\text{ref}}\mathbf{S}\right)^{-1}\mathbf{S}$ will not generally be Hermitian~\cite{symmetric_issue,symmetric_book}, which could result in decreased numerical efficiency.  The most straightforward method of ensuring that the transformed Hamiltonian is Hermitian is to pre-multiply by the overlap matrix, giving:
\begin{equation}
\mathbf{S}\left(\mathbf{H}-E_{\text{ref}}\mathbf{S}\right)^{-1}\mathbf{S}\mathbf{x}
=\left(\epsilon-E_{\text{ref}}\right)^{-1}\mathbf{S}\mathbf{x} .
\end{equation}
For this case, the eigenvalues of the original matrix will be calculated in descending order, starting from the reference energy, as demonstrated in Fig.~\ref{fig:three_methods}, which contains a diagram showing the transformation of a set of example eigenvalues following the application of shift invert.  In order to correctly calculate the conduction states, the reference energy should therefore be set between the highest required conduction band and the state immediately above (shift invert variant +).  One way to avoid this problem is to multiply the new Hamiltonian by minus one, reversing the order of calculation and therefore allowing the conduction states to be calculated in ascending order starting from the LUMO (lowest unoccupied molecular orbital), simply by setting the reference energy to be just above the HOMO (highest occupied molecular orbital) (shift invert variant -).

The shift invert method can suffer from stability problems, which can be reduced by adding an imaginary component, $\text{i}\mu$, to the reference energy, however this means that the Hamiltonian once again loses its Hermiticity, creating the possibility of imaginary eigenvalues.  This can be avoided by combining two shift invert transformations, such that a small positive imaginary component is added to the reference energy for the first transformation and a negative component is added to the second, thereby eliminating all imaginary components.  This gives the final generalized eigenvalue equation:
\begin{eqnarray}
\mathbf{S}\left[\mathbf{H}\mathbf{S}^{-1}\mathbf{H}-2E_{\text{ref}}\mathbf{H}+\left(E_{\text{ref}}^2+\mu^2\right)\mathbf{S}\right]^{-1}\mathbf{S}\mathbf{x}\\
=\left(\epsilon-E_{\text{ref}}\right)^{-2}\mathbf{x}\nonumber.
\end{eqnarray}
In this case the eigenvalues appear in an unfavourable order, such that as the transformed eigenvalues increase in energy, $|\epsilon-E_{\text{ref}}|$ decreases , i.e.\ the eigenvalues furthest from $E_{\text{ref}}$ will be found first.  Multiplying the Hamiltonian by minus one will reverse the order, returning to the situation where eigenvalues closest to the reference energy are found first (shift invert variant i).  This resembles the folded spectrum method in that the conduction and valence states again become mixed, and so a careful choice of reference energy is needed.

\paragraph{Projection} \label{subsection:projectors} 

The density operator is defined according to Eq.~(\ref{eq:dens_op}), where the $f_n$ are the occupation numbers which are assumed to be $1$ for valence states and $0$ for conduction states within the test program.  The density operator $\hat{\rho}$ is a projection operator onto the subspace of states occupied by the valence states, so that projecting $\hat{\rho}$ onto $\hat{H}$ and solving the new eigenvalue equation will give only the valence eigenstates. Alternatively, projecting with $1-\hat{\rho}$, where the 1 is defined in the psinc basis, will leave only contributions from the conduction states.  This is illustrated in Fig.~\ref{fig:three_methods}, which contains a schematic demonstrating the effect of projecting the Hamiltonian in this manner on a set of example eigenvalues. 

One problem which can arise due to the imposition of localization constraints during a calculation is that $\hat{H}$ and $\hat{\rho}$ may not commute exactly, which will result in the projected Hamiltonian no longer being Hermitian.  This can be overcome by projecting twice, so that the expression
\begin{equation}
\hat{H}-\hat{\rho}\hat{H}\hat{\rho}
\end{equation}
is used to form the new projected Hamiltonian.  However, projecting the Hamiltonian in this manner leads to an energy spectrum where all the valence energies are equal to zero, which is only desirable when all the conduction energies are negative and so more favourable in energy than the zeroed valence states.  To avoid this problem the energy spectrum is shifted so that all the valence states become higher in energy than the conduction states.  This shift must be greater than or equal to the highest conduction energy, the value of which can be easily found using conjugate gradients, as only the highest energy is required.  The projected Hamiltonian can be modified to include the shift, $\sigma$, so that the final operator is:
\begin{equation}
\hat{H}-\hat{\rho}\left(\hat{H}-\sigma\right)\hat{\rho}.
\end{equation}
In practice, the shift $\sigma$ is set to be higher than the highest conduction energy, so that in general it remains constant even when there are changes in the highest eigenvalue, adding stability to the minimization process.  If necessary, it can also be updated during the calculation.

\subsection{Results and discussion}

These five methods were tested and compared for a system with a Kronig-Penney potential~\cite{kittel} using the block update preconditioned conjugate gradients method~\cite{mostofi_prec}.  By applying the appropriate level of preconditioning and selecting a good choice of reference energies, the results in Table~\ref{table:cond_res1} were obtained.  No shift was applied for the projection method.  In attempting to choose good values for the reference energies, it was verified that a poor choice can result in significantly slower convergence.  For all of the methods the total conduction energies calculated were accurate to within $10^{-10}$ Ha of the correct result.

\begin{table}[!h] 
\caption[Comparison of conduction state methods]{Results for the different conduction methods, showing averages for time taken and the number of iterations for a total of 100 calculations with randomly generated starting guesses for the eigenvectors.  Shift invert +, - and i refer to the three variants of the shift invert method discussed in the text.  The first form of the projection method was applied, without the use of a shift.}  \label{table:cond_res1}
\begin{ruledtabular}
\begin{tabular}{lcc}
Method & Avg. time & Avg. number \\
 & taken (s) & of iterations \\
\hline
Folded spectrum & 2.39 & 182 \\
Shift invert + & 2.34 & 158 \\
Shift invert - & 2.23 & 170 \\
Shift invert i & 5.48 & 463 \\
Projection & 1.21 & 36 \\
\end{tabular}
\end{ruledtabular}
\end{table}

The results show that the different methods are fairly similar in terms of both speed and accuracy, with the projection method as the clear favorite.  An important requirement of the selected method is the need for linear-scaling.  Whilst this is hard to test within this basic implementation due to the lack of localization and sparse matrix multiplication, it can be shown that with the appropriate level of preconditioning, the number of iterations required for increasing system size remains approximately constant for the projection method.  Combined with the fact that the method mainly consists of matrix multiplications, it seems likely that favorable scaling could be achieved when implemented within local-orbital methods.

The reason for the relatively large number of iterations required for the folded spectrum method can be seen by considering the condition number, which will be higher for the folded Hamiltonian.  Using the approximate expression~\cite{goedecker19}:
\begin{equation}
\kappa \approx \frac{\left(\epsilon_{\text{max}}-\epsilon_{\text{min}}\right)}{\epsilon_{\text{gap}}},
\end{equation}
it is clear that the largest eigenvalue $\epsilon_{\text{max}}$ will be much bigger for the transformed Hamiltonian, and thus so will the condition number, $\kappa$.  Therefore  when using an iterative minimization scheme, convergence will be slower compared to solving the original equation.   

For both the shift invert and folded spectrum methods, the choice of reference energy is particularly important.  For the folded spectrum method, for example, if it is too low then unwanted valence states will be re-calculated, if it is too high then unwanted high energy conduction states will need to be calculated in order to get the lowest conduction states.  Additionally a poor choice of reference energy will result in slower convergence for the shift invert method.  For example, if the reference energy is too close to a given eigenvalue, such that the difference between $E_{\text{ref}}$ and $\epsilon$ is very small compared to the distance to other eigenvalues, the magnitude of the eigenvalue for the new system will be much greater than all other eigenvalues.  This will result in a high condition number, so care must be taken to find a good reference energy.  The projection method, however, has the advantage that no reference energy is required and therefore it is more automatic.  Additionally, the density matrix is already calculated within a local-orbital calculation and so can easily be reused.

For the case of all three methods, the accuracy of the conduction states will clearly be affected by the accuracy with which the potential has been calculated.  However, the projection method will also be affected by the accuracy of the valence density matrix, whereas the folded spectrum and shift invert methods will not.  This will be particularly significant when the localization and truncation approximations required for linear-scaling behaviour are applied.

\subsection{Implementation in ONETEP}

The methods outlined above were applied directly to the solution of an eigenvalue equation.  However in a real \textsc{onetep} calculation, the system is solved using a density matrix scheme, within the representation of a basis of NGWFs.  It is therefore necessary to adapt the methods described above for use within this context.  As the projection method has proven to be the most favorable, this is the one which was subsequently focussed on.

Two sets of NGWFs are now required, $\{|\phi_{\alpha}\rangle\}$ for the valence states, and $\{|\chi_{\alpha}\rangle\}$ for the conduction states.  The ground state \textsc{onetep} calculation already provides access to the valence density matrix ${\bm{\rho}}$ and kernel $\mathbf{K}$, overlap matrix $\mathbf{S_{\phi}}$ and Hamiltonian $\mathbf{H_{\phi}}$.  The additional conduction matrices will be labelled as follows: $\mathbf{S_{\chi}}$ is the conduction overlap matrix, $\mathbf{T}$ is the valence-conduction cross overlap matrix defined as $T_{\alpha\beta}=\langle \phi_{\alpha} | \chi_{\beta} \rangle$, $\mathbf{H_{\chi}}$ is the (unprojected) conduction Hamiltonian, $\mathbf{H_\chi^{\textrm{proj}}}$ is the projected conduction Hamiltonian, $\mathbf{Q}$ is the conduction density matrix and $\mathbf{M}$ is the conduction density kernel.  These are all represented by atom-blocked sparse matrices~\cite{onetep3,onetep_sparse}, such that all matrix-matrix operations are possible in asymptotically linear-scaling computational effort, due to the strict truncation.

The final expression for the projected conduction Hamiltonian, including the shift, $\sigma$, is therefore defined as follows:
\begin{eqnarray}
\left(H_\chi^{\textrm{proj}}\right)_{\alpha\beta}&=&\langle \chi_\alpha|\hat{H}-\hat{\rho}\left(\hat{H}-\sigma\right)\hat{\rho}|\chi_\beta\rangle\\ \nonumber
&=&\left(H_\chi\right)_{\alpha\beta} -\left(T^\dag K H_\phi KT\right)_{\alpha\beta}\\ \nonumber
&&+\sigma\left(T^\dag K S_\phi KT\right)_{\alpha\beta}
\end{eqnarray}
The energy expression $E=\text{tr}\left[\mathbf{M}\mathbf{H_\chi^{\textrm{proj}}}\right]$ can then be minimized by optimizing both the set of conduction NGWFs and the conduction kernel.  Extra terms will be needed in the NGWF gradient, but otherwise this follows the same procedure as a standard \textsc{onetep} calculation, without the need for self-consistency.  Once the set of conduction NGWFs has been optimized, the Hamiltonian can be diagonalized in a joint basis of valence and conduction NGWFs to give an improved eigenvalue spectrum.  This allows eigenvalues and other properties to be calculated in a basis that is capable of representing both the valence and conduction states of the system.

\subsection{Calculating optical spectra}

As stated in the introduction, the calculation of experimental spectra in general and optical spectra in particular can be highly useful both in predicting and understanding experimental results and can be applied to a diverse range of problems.  The method followed for the calculation of optical absorption spectra is that~\cite{pickard72} applied in \textsc{castep}~\cite{castep44}, a cubic-scaling plane-wave pseudopotential (PWPP) DFT code which can use the same pseudopotentials as \textsc{onetep}, and so is ideal for comparison of results.  The method employed is described briefly below.

Starting from time-dependent perturbation theory, one can derive Fermi's golden rule, an expression giving the probability of a particular electronic transition.  It involves a joint density of states between valence and conduction states, which is weighted by optical matrix elements.  Matrix elements with a value of zero indicate that a given transition is forbidden, whereas nonzero matrix elements define the strength of the transition.  These matrix elements take the form of a complex exponential, which in the long-wavelength limit can be related to position matrix elements using the dipole approximation, where the exponential is expanded in a Taylor series and terms above first order are neglected.  In this manner, the imaginary part of the dielectric function can be written as:
\begin{equation} \label{eq:imag_diel}
\varepsilon_2\left(\omega\right)=\frac{2e^2\pi}{\Omega\varepsilon_0}\sum_{\mathbf{k},v,c}\left|\langle\psi_{\mathbf{k}}^{c}|\mathbf{\hat{q}}\cdot\mathbf{r}|\psi_{\mathbf{k}}^{v}\rangle\right|^2\delta\left(E_{\mathbf{k}}^{c}-E_{\mathbf{k}}^{v}-\hbar\omega\right) ,
\end{equation}
where $v$ and $c$ denote valence and conduction bands respectively, $|\psi_{\mathbf{k}}^{n}\rangle$ is the $n$th eigenstate at a given $\mathbf{k}$-point with a corresponding energy $E_{\mathbf{k}}^n$, $\Omega$ is the cell volume, $\mathbf{\hat{q}}$ is the direction of polarization of the photon and $\hbar\omega$ its energy.  In principle this includes a $\mathbf{k}$-point sum over the entire Brillouin zone, however as with ground-state \textsc{onetep} calculations, it is assumed that a large enough supercell will be used such that only the gamma point need be considered.  As the system size increases, this will become an increasingly exact approximation, so that the accuracy of the density of states will improve for bigger systems.  This could be extended in future using methods for interpolating band structures in \textsc{onetep} that will be published elsewhere.  For the purposes of this work, however, all calculations have been restricted to the gamma point only.  From the imaginary part of the dielectric function one can then also calculate the real part using the appropriate Kramers-Kronig relation.

In both \textsc{onetep} and \textsc{castep} periodic boundary conditions are used, in which the position operator is known to be undefined.  Due to the strict localization of the NGWFs in \textsc{onetep}, it is possible to calculate position matrix elements between eigenstates for molecules, providing the NGWF radii are sufficiently small such that no NGWFs associated with the molecule overlap with any NGWFs associated with its periodic image.  However for periodic systems it becomes necessary to use the momentum operator.  Momentum matrix elements can be easily related to position matrix elements by considering the commutator with the Hamiltonian, but when non-local pseudopotentials are being used, one must be careful to include the commutator between the position operator and the non-local potential.  The relation is thus written~\cite{read_needs}:
\begin{equation}
\langle\phi_f|\mathbf{r}|\phi_i\rangle = \frac{1}{\text{i}\omega m}\langle\phi_f|\mathbf{p}|\phi_i\rangle + \frac{1}{\hbar\omega}\langle\phi_f|\left[\hat{V}_{\text{nl}},\mathbf{r}\right]|\phi_i\rangle .
\end{equation}
In practice, the commutator term is calculated using the following identity~\cite{nonloc_com_deriv}:
\begin{eqnarray}
&&\left(\nabla_\mathbf{k}+\nabla_\mathbf{k'}\right)\left[\int e^{-\text{i}\mathbf{k}\cdot\mathbf{r}} V_{\text{nl}}\left(\mathbf{r},\mathbf{r'}\right) e^{\text{i}\mathbf{k'}\cdot\mathbf{r'}} \text{d}\mathbf{r}\ \text{d}\mathbf{r'}\right] \\
&=&i\int e^{-\text{i}\mathbf{k}\cdot\mathbf{r}}\left[V_{\text{nl}}\left(\mathbf{r},\mathbf{r'}\right)\mathbf{r'}-\mathbf{r}V_{\text{nl}}\left(\mathbf{r},\mathbf{r'}\right)\right] e^{\text{i}\mathbf{k'}\cdot\mathbf{r'}} \text{d}\mathbf{r}\ \text{d}\mathbf{r'} \nonumber,
\end{eqnarray}
where the derivative can either be calculated directly or using finite differences in reciprocal space.  The matrix elements are thus calculated in this manner and used to form a weighted density of states, which is smeared using Gaussian functions.  

For the purposes of comparison with experiment, it is sometimes desirable to make use of the scissor operator, whereby the conduction band energies are rigidly shifted upwards such that the DFT Kohn-Sham band gap is equal to experimental values.  Whilst this is not an \emph{ab initio} correction, in practice relatively good agreement can be found with experiment in this manner for many systems without the need for more computationally intensive methods, such as the \emph{GW} approximation, although there will be a number of occasions when it becomes necessary to use less approximate methods.

\section{Results and discussion} \label{sec:results}

\subsection{Metal-free phthalocyanine}\label{sec:phthalo_results}

As stated in section~\ref{sec:lin_sc_dft}, \textsc{onetep} is particularly efficient at treating molecules, and so metal-free phthalocyanine was chosen as a good test system  on which to apply the conduction state method.  As it contains only 58 atoms, calculations could also be performed using \textsc{castep}, which has been used for all the traditional PWPP DFT results given.  Corresponding plane-wave/psinc kinetic energy cut-offs and identical norm-conserving pseudopotentials were used for both codes.  In all calculations the local-density approximation (LDA) exchange-correlation functional was used.  Phthalocyanines and their derivatives are commonly used as dyes and are also of interest in a number of other fields, including use in photovoltaic cells~\cite{kobay3} and molecular spintronics~\cite{shen_pc_spin} and so metal-free phthalocyanine also provides an interesting test case for the calculation of optical absorption spectra.

\begin{figure}[]
    \begin{center}
    \includegraphics[scale=0.3]{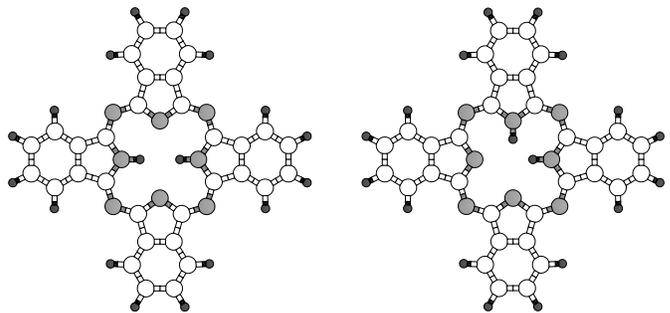}
  \caption{Schematics showing the structures of the \emph{trans} (left) and \emph{cis} (right) isomers of metal-free phthalocyanine.  C atoms are shown in white, N atoms in gray and H atoms in black.}
  \label{fig:pc_strucs}
  \end{center}
\end{figure}

\begin{figure}[]
    \begin{center}
    \includegraphics[scale=0.58, angle=270]{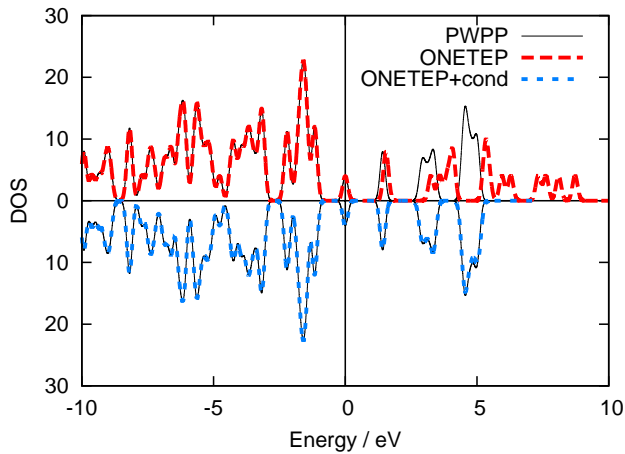}
  \caption{(Color online) Density of states comparing results from \textsc{onetep} with PWPP results for the geometry-optimized structure of metal-free phthalocyanine, plotted with a Gaussian smearing width of 0.1~eV, using conduction NGWF radii of 16~Bohr.  The DOS is truncated after the first 16 conduction states, and the \textsc{onetep}+cond curve shown is calculated in the joint valence-conduction NGWF basis.}
  \label{fig:pc_dos_cond}
  \end{center}
\end{figure}

The atomic coordinates for metal-free phthalocyanine were taken from neutron diffraction data~\cite{hoskins6} with $\mathrm{C}_{2h}$ symmetry and the inner H atoms attached to opposite N atoms.  Additional symmetry constraints were then applied by averaging the atomic positions to give the higher symmetry $\mathrm{D}_{2h}$ with the inner H atoms attached to both opposite and adjacent N atoms (\emph{trans} and \emph{cis} forms respectively), and finally a geometry optimized structure was calculated using traditional DFT, which also has a \emph{trans}-$\mathrm{D}_{2h}$ symmetry but differs in bond lengths from the other $\mathrm{D}_{2h}$ structure.  Diagrams of the \emph{trans} and \emph{cis} forms are shown in Fig.~\ref{fig:pc_strucs}.  Table~\ref{table:pc_struc_energies} shows the ground state energies for each structure relative to the geometry optimized result, with the higher symmetry structures lower in energy.  Very good agreement is achieved between the \textsc{onetep} and traditional DFT results.  Both the \textsc{onetep} and PWPP calculations were performed at a kinetic-energy cut-off of 1046~eV, with the \textsc{onetep} valence NGWFs at a fixed radius of 12~Bohr, with one NGWF per H atom, and four each per C and N atom.  Sixteen conduction states were optimized, with four conduction NGWFs for each atomic species, and a radius of 16~Bohr was used for the density of states (DOS) calculations, whilst 13~Bohr was sufficient to achieve almost perfect agreement with traditional DFT for the optical absorption spectra.  This difference in NGWF radii required for good convergence of DOS and optical absorption spectra is discussed in section~\ref{sec:limits}.  The DOS for the geometry optimized structure is shown in Fig.~\ref{fig:pc_dos_cond}, which compares \textsc{onetep} results both with and without conduction NGWFs to those found using the PWPP method.  Without the conduction NGWFs, the \textsc{onetep} results differ greatly from the PWPP results, but with the addition of conduction NGWFs, excellent agreement with the PWPP method is achieved.  A state-by-state comparison confirmed the existence of a one-to-one correspondence between the \textsc{onetep} and \textsc{castep} conduction eigenstates.

\begin{table}[] 
\caption[]{Comparison between \textsc{onetep} and PWPP ground-state total energies for four different structures of metal-free phthalocyanine, relative to the lowest energy geometry optimized structure.  The energy difference between the \textsc{onetep} and PWPP results for the geometry optimized structure is 0.163~eV.}  \label{table:pc_struc_energies}
\begin{ruledtabular}
\begin{tabular}{lcc}
\multirow{2}{*}{Structure} & \multicolumn{2}{c}{$E-E_{\text{geom}}$ / eV} \\
& \textsc{onetep} & PWPP \\ \hline 
$\mathrm{C}_{2h}$ & 1.554 & 1.553  \\
\emph{cis}-$\mathrm{D}_{2h}$ & 1.875 & 1.874 \\
\emph{trans}-$\mathrm{D}_{2h}$ & 0.952 & 0.951 \\
\end{tabular}
\end{ruledtabular}
\end{table}

Optical absorption spectra were then calculated using both the position operator and the momentum operator (including the non-local commutator) for all four structures, and in all cases the two methods agreed almost perfectly with the PWPP results for the energy range considered.  The addition of a greater number of conduction states is unnecessary for this energy range, confirming that the calculation of unbound conduction states will not always be needed.

\begin{figure}[]
    \begin{center}
    \includegraphics[scale=0.58, angle=270]{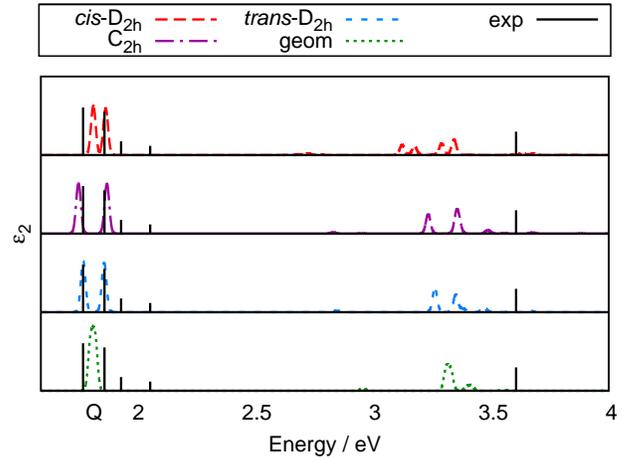}
  \caption{(Color online) The imaginary component of the dielectric function calculated in \textsc{onetep} and plotted for four different structures of metal-free phthalocyanine as indicated on the graph (`geom' refers to the geometry optimized structure).  Results from the PWPP method are indistinguishable and so not plotted.  A Gaussian smearing width of 0.01~eV is used, with conduction NGWF radii of 13~Bohr and a scissor operator of 0.4~eV.  Experimental results in solution (labelled `exp') are also included with the peaks shown as vertical lines for clarity and the calculated results vertically scaled arbitrarily for easier comparison.  The position of the Q-bands is indicated below the x-axis.}
  \label{fig:pc_spec_cond}
  \end{center}
\end{figure}

It should be emphasized here that the aim of this work is calculate absorption spectra within DFT and so find good agreement with conventional DFT implementations, rather than go beyond DFT and achieve good agreement with experiment.  However, useful insight can be achieved through comparision with experiment, and so the absorption spectra for the four structures were compared with experimental results in solution~\cite{kobay3}, applying a scissor operator of 0.4~eV, and arbitrarily scaling the height of the imaginary part of the dielectric function to facilitate easier comparison with experiment, as shown in Fig.~\ref{fig:pc_spec_cond}.  The spectra are indeed distinguishable, despite the very small differences in the atomic structures.   

It is also possible to identify the transitions responsible for the peaks, with the split Q-band peaks (indicated in Fig.~\ref{fig:pc_spec_cond}) being due to HOMO-LUMO and HOMO-LUMO+1 transitions and the degree of splitting within the peak therefore due to the energy difference between the LUMO and LUMO+1 bands.  It is accepted that the lower symmetry of the metal-free phthalocyanine structure as compared to metal phthalocyanines is the cause for this Q-band splitting, which is not observed for metal phthalocyanines.  This agrees with the observation that the higher symmetry \emph{trans}-$\mathrm{D}_{2h}$ structure exhibits a lower degree of splitting than the \emph{trans}-$\mathrm{C}_{2h}$ structure.  The Q-band splitting for the geometry optimized structure is 0.02~eV, which is significantly less than the experimental value of 0.09~eV, implying that the LDA is not sufficiently accurate to calculate the correct structure.

There have already been a number of studies~\cite{fukuda4,Gong11,cortina1,day7} of the electronic structure and absorption spectra of metal-free phthalocyanine, with which the above results are consistent, confirming that this is a useful system to demonstrate the ability of theoretical optical absorption spectra as implemented here to distinguish between similar geometries.

\subsection{Poly(\emph{para}-phenylene)}

\begin{figure}[]
    \begin{center}
    \includegraphics[scale=0.25]{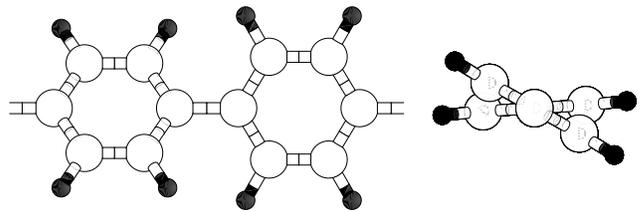}
  \caption{Schematic showing the structure of a unit of poly(\emph{para}-phenylene) from two directions.  C atoms are shown in white and H atoms in black.}
  \label{fig:ppp_struc}
  \end{center}
\end{figure}

\begin{figure}[]
    \begin{center}
    \includegraphics[scale=0.58, angle=270]{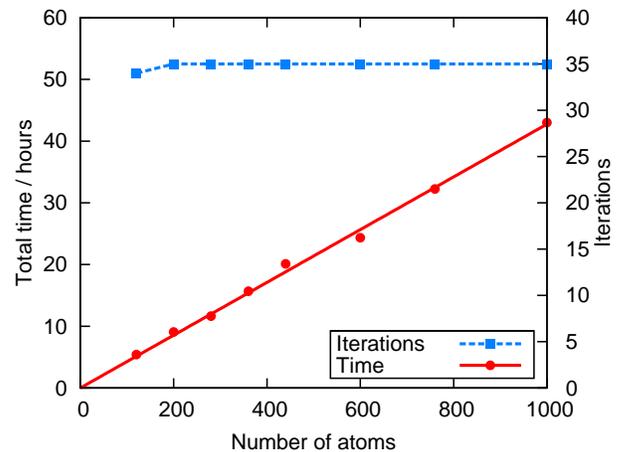}
  \caption{(Color online) Graph showing the scaling of \textsc{onetep} conduction calculations for increasing chain lengths of PPP.  Calculations were performed on 8 nodes and therefore a total of 32 cores.  The total time taken for \textsc{onetep} is approximately linear up to 1000 atoms and the number of NGWF iterations required for convergence shown in the inset is shown to be constant with an increasing number of atoms.}
  \label{fig:ppp_scaling_cond}
  \end{center}
\end{figure}

Conjugated polymers such as poly(\emph{para}-phenylene) have a wide range of applications due to their electroluminescent properties, including LEDs and solar cells~\cite{ppp_mol,ppp_lane,ppp_burr} and so this also provides an interesting system to study as a test case for the calculation of optical absorption spectra.  As a periodic system, it is also ideal for testing the scaling of the projection method, by increasing the size of the unit cell and comparing the time taken to calculate the conduction states.  The structure for two unit cells was obtained by performing a geometry optimization with a PWPP code using the structure of Ambrosch-Draxl \emph{et al}.~\cite{ppp_struc} as a starting guess, with the final structure shown in Fig.~\ref{fig:ppp_struc}.  A cut-off energy of 1115~eV was found to be necessary for good convergence of the results.  All calculations were performed at the Gamma point only, with no $\mathbf{k}$-point sampling, to allow for direct comparison between the two codes.  Ground state calculations were first performed with one NGWF per H atom and four NGWFs per C atom and a fixed radius of 10~Bohr.  Conduction calculations were then performed using four NGWFs for all atomic species with a fixed radius of 14~Bohr.  The number of conduction states calculated was set to include all negative eigenvalues for the smallest system (corresponding to two unit cells of PPP) and increased linearly with system size.  Fig.~\ref{fig:ppp_scaling_cond} shows the scaling behavior of \textsc{onetep} for the conduction calculation.  Neither the valence nor conduction density kernels were truncated, however, the behaviour of \textsc{onetep} is shown to be approximately linear up to 1000 atoms, and it is expected that this trend will continue up to larger system sizes. 

\begin{figure}[]
    \begin{center}
    \includegraphics[scale=0.58, angle=270]{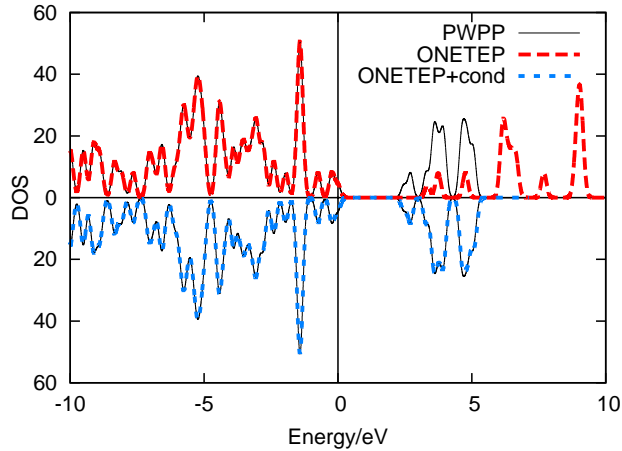}
  \caption{(Color online) Density of states calculated using \textsc{onetep} and a PWPP code for 120 atoms of PPP with conduction NGWF radii of 14~Bohr and a Gaussian smearing width of 0.1~eV.  The \textsc{onetep}+cond curve is from the joint valence-conduction NGWF basis, and has been plotted so that only those conduction states which have been optimized are included.  The same number of states have been plotted for both the \textsc{onetep} and PWPP curves.}
  \label{fig:ppp_dos_cond}
  \end{center}
\end{figure}

\begin{figure}[]
    \begin{center}
    \includegraphics[scale=0.58, angle=270]{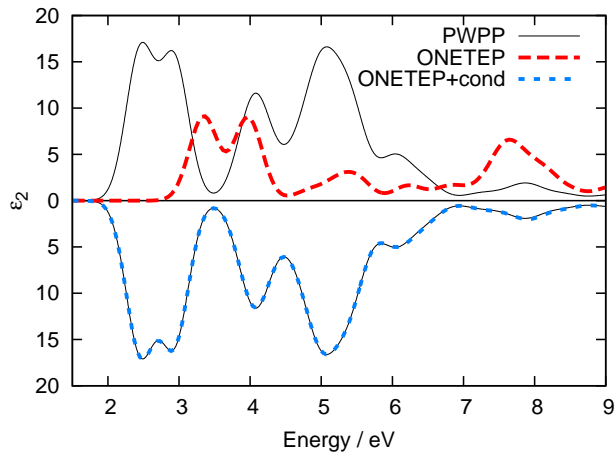}
  \caption{(Color online) The imaginary component of the dielectric function calculated using a traditional PWPP code and \textsc{onetep} both with and without a conduction calculation for 120 atoms of PPP.  Conduction NGWF radii of 14~Bohr and a Gaussian smearing width of 0.2~eV are used.}
  \label{fig:ppp_spec}
  \end{center}
\end{figure}

The density of states was plotted for varying chain lengths of PPP, with the graph for 120 atoms shown in Fig.~\ref{fig:ppp_dos_cond}.  As with metal-free phthalocyanine, excellent agreement with the PWPP results is achieved for the conduction calculation.  The imaginary component of the dielectric function was also calculated for varying chain lengths, using the momentum operator formulation.  The result for 120 atoms is shown in Fig.~\ref{fig:ppp_spec}.  Again, nearly perfect agreement with the PWPP method was achieved with the conduction NGWF basis, whereas the valence NGWF basis only calculation showed big discrepancies not only in the positions of the peaks, but also in the relative strengths.

\subsection{Limitations of the method}\label{sec:limits}

The projection method has proven to be a good method of optimizing a set of NGWFs that are capable of representing the conduction states to a good degree of accuracy.  However, there are some limitations to the method, which will be discussed below.

One limitation which cannot be overcome is the inability to represent completely delocalized and unbound states, which is to be expected with a localized basis.  With increasing NGWF radii the eigenvalues tend towards the correct Kohn-Sham eigenvalues, however when one uses such large radii the prefactor of the calculation becomes dominant, so that even though the overall behaviour is still linear-scaling, the crossover point at which the method becomes quicker than cubic-scaling codes will occur at systems with a greater number of atoms.  However, for applications considered here, notably the calculation of optical absorption spectra, often only lower energy bound states are required, as many of the interesting features in optical spectra are transitions between bands close to the gap and one is interested in a relatively low energy range.  Therefore in practice this limitation on the method is less serious than it first appears to be.  Additionally, it has been observed that the lower energy conduction states converge with respect to NGWF radius faster than those with higher energy, and so if the lower energy bound states only are considered, it no longer becomes necessary to use such large NGWF radii to achieve a good level of convergence in the optical absorption spectra.  It was for this reason that smaller conduction NGWF radii were used for the absorption spectra compared to the DOS of metal-free phthalocyanine, as presented in section~\ref{sec:phthalo_results}.

\begin{figure}[]
    \begin{center}
    \includegraphics[scale=0.58, angle=270]{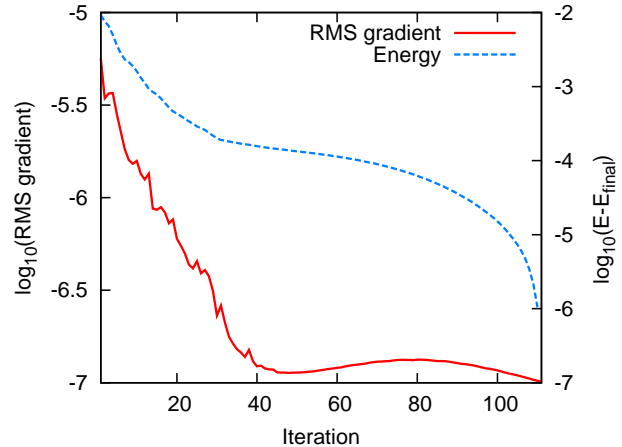}
  \caption{(Color online) Demonstration of the appearance of a local minimum, where the RMS gradient increases for a period whilst the energy continues to decrease.  This was for the geometry-optimized structure of metal-free phthalocyanine at a radius of 18~Bohr with four extra states being optimized.}
  \label{fig:loc_minb}
  \end{center}
\end{figure}

\begin{figure}[]
    \begin{center}
    \includegraphics[scale=0.58, angle=270]{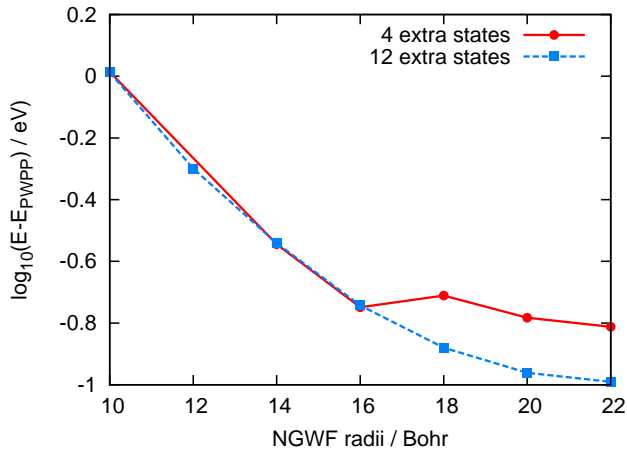}
  \caption{(Color online) Comparison of conduction energy convergence with respect to conduction NGWF radii for different numbers of extra states, for the geometry-optimized structure of metal-free phthalocyanine.  The energy difference is calculated with respect to the traditional PWPP result.  A discontinuity appears in the curve at 18~Bohr when four additional states are optimized, demonstrating problems with local minima, whilst optimizing 12 additional states is sufficient to overcome the problem.}
  \label{fig:loc_mina}
  \end{center}
\end{figure}

It has also been observed that it is sometimes possible to become trapped in a local minimum when optimizing the conduction orbitals.  This behaviour is characterized by slow convergence of the conduction NGWFs, wherein the RMS gradient stagnates or increases while the energy continues to decrease; or by sharp jumps in the energy with increasing conduction NGWF radii, rather than the expected smooth convergence.  Examples of both of these features can be seen in Figs.~\ref{fig:loc_minb} and~\ref{fig:loc_mina} respectively.  This behaviour has been seen to occur due to an unfavourable ordering of the energy eigenstates in the unoptimized basis of NGWFs, so that the NGWFs are optimized for some eigenstates which will eventually be higher in energy at the expense of those which will eventually be lower in energy.  This behaviour is strongly system dependent, however it can be overcome by initially optimizing a greater number of conduction states than required, then reducing the number of states to that actually required, regenerating the conduction density kernel and proceeding with the calculation.  This first stage aims to overcome the problem of poor initial ordering of states, whilst the second stage will allow for closer optimization of those states actually required.   This is illustrated by Table~\ref{table:pc_order_state}, where the LUMO+14 state is initially much higher in energy and so if no additional states are included the NGWFs are not optimized to represent it, so that it ends up significantly higher in energy than other states.  If, however, four additional states are included, this is sufficient to reorder the states and it becomes lower in energy.

\begin{table}[!h] 
\caption[]{Initial energies and values after 5 iterations both with and without optimizing extra states for three different eigenstates of the geometry-optimized structure of metal-free phthalocyanine with conduction NGWF radii of 14~Bohr.  States shown in bold are those which are among the 16 lowest states, and thus being included in the conduction NGWF optimization.  Without the optimization of extra states, the LUMO+14 state is not optimized and so remains high in energy, whilst with the addition of four extra states, the correct order is found and the LUMO+14 is significantly lowered in energy.}  \label{table:pc_order_state}
\begin{ruledtabular}
\begin{tabular}{lccc}
State & Initial & 0 extra states & 4 extra states \\
\hline
LUMO+14 & 0.628 & $>$0.368 & \bf{-0.042} \\
LUMO+15 & \bf{0.355} & \bf{0.045} & \bf{0.039} \\
LUMO+16 & \bf{0.259} & \bf{0.082} & 0.061 \\
\end{tabular}
\end{ruledtabular}
\end{table}

As well as the above-mentioned problems, there are a number of parameters which require more careful consideration when selecting appropriate values than in a ground state \textsc{onetep} calculation, where they can be set automatically.  This includes the number of conduction states one is trying to represent, the number of NGWFs one chooses for each atom, the number of additional states to be optimized and the number of iterations for which these extra states are optimized.  Some of these parameters, such as the number of iterations to perform in the first stage of the local minima avoiding scheme, have less of an effect on the final result, but for many of these parameters, the effect of different values appears to be strongly system-dependent.  One must therefore perform careful convergence tests to ensure that the resulting states do not correspond to any local minima.  This will require variation of the number of NGWFs per atom, convergence with respect to NGWF radii, and an increase in the number of extra conduction states requested, until consistent results are achieved, with a smooth curve of energy against NGWF radii, and sensible convergence of the NGWFs during a calculation.  By following these strategies one can become confident that accurate results have been achieved.

It should also be noted that the iterative energy minimization scheme used here requires the presence of a band gap, which for the conduction calculation translates as a gap between the highest optimized conduction state, and the lowest unoptimized conduction state.  As one approaches the continuum of conduction states, this gap will become increasingly small, which could result in poor convergence behaviour.  

Finally, it is observed that whilst problems have been encountered with the projection method, a clear strategy has been outlined both for identifying and resolving them. 

\section{Conclusions}
\label{sec:conclusions}

In conclusion, a methodology has been presented for the accurate calculation of the unoccupied Kohn-Sham states within a linear-scaling DFT context.  Excellent agreement was achieved with traditional PWPP results for lower lying conduction states, although the use of localized basis functions is not ideal for higher-energy delocalized conduction states.  Additionally, a strategy has been outlined for both identifying and avoiding the problem of local minima which have been seen to occur.

The existence of a localized basis set capable of representing the Kohn-Sham conduction states in \textsc{onetep} has enabled the calculation of optical absorption spectra using Fermi's golden rule as a first approximation.  This methodology allows one to take advantage of large-scale linear-scaling calculations and extract useful information, which can be compared to experimental results and aid with the interpretation of those results.  In particular, the ability to identify the transitions responsible for a given peak and compare spectra from very similar atomic structures has been demonstrated, through the application to both a molecular and an extended system.  Furthermore, it also forms the basis of future extensions both to more accurate methods of calculating optical spectra, and to calculating other types of spectra, such as electron energy loss spectra and x-ray absorption and photoemission spectra. 

\begin{acknowledgments}
This work was supported by the UK Engineering and Physical Sciences Research Council (EPSRC).  Calculations were performed on CX1 (Imperial College London High Performance Computing Service).  N.D.M.H. acknowledges the support of the Engineering and Physical Sciences Research Council (EPSRC Grant No. EP/G055882/1) for postdoctoral funding through the HPC Software Development call 2008/2009. P.D.H. acknowledges support from the Royal Society in the form of a University Research Fellowship.

\end{acknowledgments}

\end{document}